\documentclass[hyper]{JHEP3} 

\usepackage{graphicx}
\usepackage{epsfig}
\usepackage{cite}
\usepackage{amsmath}

\makeatletter

\newcommand{\fmslash}[2][0mu]{%
  \mathchoice
    {\fmsl@sh\displaystyle{#1}{#2}}%
    {\fmsl@sh\textstyle{#1}{#2}}%
    {\fmsl@sh\scriptstyle{#1}{#2}}%
    {\fmsl@sh\scriptscriptstyle{#1}{#2}}}
\newcommand{\fmsl@sh}[3]{%
  \m@th\ooalign{$\hfil#1\mkern#2/\hfil$\crcr$#1#3$}}
\makeatother

\newcommand{\met}{\not \!\! E_T}

\newcommand{\mptvec}{\not \!\! \vec{P}_T}
\newcommand{\mblmax}{$m_{b\ell}^{max}$}

\title{\Large Revisiting Combinatorial Ambiguities at Hadron Colliders with $M_{T2}$}

\author{Philip Baringer, Kyoungchul Kong, Mathew McCaskey, Daniel Noonan \\
        Department of Physics and Astronomy,  
        University of Kansas, Lawrence, KS 66045 USA \\
        E-mail: \email{baringer@ku.edu, kckong@ku.edu, mccaskey@ku.edu, dnoonan@ku.edu}
        }

\received{\today}               
\accepted{\today}               

\preprint{ 
          \today 
          } 

\abstract{
We present a method to resolve combinatorial issues in multi-particle final states at hadron colliders. 
The use of kinematic variables such as $M_{T2}$ and invariant mass significantly reduces combinatorial ambiguities 
in the signal, but at a cost of losing statistics. 
We illustrate this idea with gluino pair production leading to 4 jets $+\met$ in the final state as well as 
$t\bar{t}$ production in the dilepton channel.
Compared to results in recent studies, our method provides greater efficiency with similar purity}

\keywords{Beyond Standard Model, Hadronic Colliders, Supersymmetry Phenomenology}

\begin{document} 

\section{Introduction}
\label{sec:intro}

The Tevatron Run II at Fermilab and the recent turn-on of the Large Hadron Collider (LHC) at CERN 
are beginning to explore the physics of the Terascale. There are sound theoretical reasons to believe that 
new physics beyond the Standard Model (BSM) is going to be revealed in those experiments.
Perhaps the most compelling phenomenological evidence for BSM particles and interactions at the TeV scale 
is provided by the dark matter problem. It is a tantalizing coincidence that a neutral, weakly interacting massive particle 
(WIMP) in the TeV range can explain all of the observed dark matter in the Universe. 
A typical WIMP does not interact in the detector and can only manifest itself as missing energy. 
The WIMP idea therefore greatly motivates the study of missing energy signatures at the Tevatron and the LHC.
Recently, new ideas of various kinematic methods for the determination of masses, spins, and couplings have attracted a lot of interest 
in missing energy signatures at colliders (see \cite{Barr:2010zj, Barr:2011xt, Barr:2011im} for recent reviews). 
They not only offer methods for the determination of masses and spins but also provide good background rejection and enhance discovery potential.

Unlike lepton colliders, the environment at hadron colliders is much more complex and often make such tasks difficult. 
One issue more prevalent in hadron colliders is the combinatorial ambiguities in events, especially in events with jets. 
In general, the event topology at hadron colliders will typically contain a number of jets. 
Some come from the decays of heavier particles in the signal and others may originate from initial state radiation (ISR).
The problem gets worse due to detector limitations such as finite resolution, inability of distinguishing quarks from anti-quarks, etc.
All these jets pose a severe combinatorics problem: 
which one of the many jets in an event is the correct one to assign to a particular decay?  

Some of the existing studies in the literature simply take for granted that the correct jet can be somehow identified.  
Others select the jet by matching to the true quark jet in the event generator output, which of course is unobservable. 
The severity of the jet combinatorics problem is rather model dependent and in practice
how well it can be dealt with depends on the individual case at hand. 
For example, if the mass spectrum is relatively broad, 
one might expect a jet from the decay of a heavier particle to be among the hardest in the event. 
This information can be used to improve the purity of the sample. Fortunately, there exists a method (the mixed event technique) 
which should, at least in principle, remove the effects from incorrect jet combinations \cite{Hinchliffe:1996iu}. 
This method has been successfully applied to measuring SUSY masses at the SPS1a study point \cite{Ozturk:2007ap} 
(see \cite{Dutta:2011gs} for a more recent study). 
However, this only works statistically and does not discriminate on an event-by-event basis. 
The existence of invisible particles adds difficulty in resolving combinatorial issues since it prohibits 
us from reconstructing the whole signal. 

Combinatorial issues appear at all different levels of an event (i.e. from ISR, decays of heavier particles, etc.).
A general solution to {\it all} these problems would be very difficult.
We may be able to achieve this statistically by relying on a likelihood analysis or 
matrix element methods \cite{Alwall:2010cq}.
However, this task requires knowledge of new physics which may not be available at the time of discovery. 
Therefore, it is desirable to find a solution that is model-independent. 
In fact, some studies on kinematic variables in the literature already indicate that these variables may be useful for mitigating combinatorial difficulties 
associated with ISR and the signal jets \cite{Alwall:2009zu,Nojiri:2010mk} 
\footnote{See \cite{Konar:2008ei, Konar:2010ma,Krohn:2011zp,Alwall:2007fs,Alwall:2010cq,Plehn:2005cq} for different approaches.}.

The ISR (rather than always proving to be an obstacle) can actually be helpful in the study of BSM physics.
This has already been demonstrated in several recent studies.  In Ref. \cite{Alwall:2008va}, 
ISR was shown to make BSM events more prominent by giving pair-produced new-physics states something to recoil against, 
thus increasing both $\displaystyle{\not} E_T$ and $H_T$. 
For the $M_{T2}$ variable (defined in Section $3.1$ below)~\cite{Lester:1999tx,Barr:2003rg}, it is known that the transverse momentum of ISR makes the kink structure 
more pronounced (see references in \cite{Barr:2010zj}). Usually ISR overestimates the expected end point, which changes depending on the size of the
transverse momentum. To make better measurements one needs to understand systematically 
the effect of ISR as a function of transverse momentum \cite{Matchev:2009iw, Burns:2008va,Konar:2009wn,Matchev:2009ad}.

In this paper, we concentrate on combinatorial ambiguities among the particles in the signal.
It is important to resolve or mitigate this combinatorial background issue.  
Solutions can be directly used to improve experimental measurements such as the forward-backward asymmetry in the top quark pair production, the $W$-helicity, 
top mass measurements, etc.
In terms of BSM, they obviously become more important since most BSM models predict a WIMP candidate which, when produced, will show up as missing transverse momentum in the detector.
An appropriate understanding of combinatorics may even be able to reduce SM backgrounds.

In resolving combinatorial ambiguities in signal events, recently Ref. \cite{Rajaraman:2010hy} argued 
that the proposed $p_T$ versus invariant mass method performs better than the well known hemisphere method \cite{Ball:2007zza}. 
This was shown for parton level events without detector effects, backgrounds or ISR. 
The results in the previous study are quoted in terms of the efficiency and purity of the samples.  For the example point taken in \cite{Rajaraman:2010hy} 
(the 3 body decay of 600 GeV gluino ($\tilde{g}$) into 2 jets and a neutralino ($\tilde{\chi}_1^0$) of 100 GeV), the authors have obtained 
91.7\% purity and 3.1\% efficiency for the diquark system with an invariant mass less than 500 GeV and a transverse momentum greater than 450 GeV. 

The purpose of this paper to revisit the proposed method in \cite{Rajaraman:2010hy} and 
improve the obtained efficiency and purity by utilizing the $M_{T2}$ variable in addition to invariant mass. 
Section~\ref{sec:pTMmethod} begins with a review of the method described in \cite{Rajaraman:2010hy}.
In Section~\ref{sec:gg} we propose our new method for resolving combinatorial ambiguities in the context of 
gluino pair production with 4 jets $+\met$ in the final state.  
Following the spirit of Ref. \cite{Rajaraman:2010hy} we analyze our method under
similar assumptions in order to make a suitable comparison.  
Then we determine the effectiveness of this method by considering different mass splittings, possible 
uncertainties in the gluino and neutralino masses, and ISR.
Next, in Section \ref{sec:ttbar}, we apply the same method to the $t\bar{t}$ dilepton system to resolve the two-fold ambiguity 
in pairing a b-tagged jet ($b$) and a lepton ($\ell$). 
Section~\ref{sec:conclusions} is reserved for our conclusions.


\section{The $p_T$ versus $M$ method}
\label{sec:pTMmethod}

In this section we briefly review the $p_T$ versus $M$ method in Ref. \cite{Rajaraman:2010hy}. 
For the discussion on combinatorial ambiguities at hadron colliders, the authors consider 
gluino pair production with each gluino decaying to the lightest neutralino and two quarks: 
$\tilde g \tilde g \to q q \bar{q} \bar{q} \tilde{\chi}_1^0 \tilde{\chi}_1^0$. 
All quarks are treated at parton level and the effects of ISR and parton showering are not considered. 
Following the example used in \cite{Rajaraman:2010hy}, we consider the off-shell squark case ($m_{\tilde{q}}  > m_{\tilde{g}}$)
with a gluino mass $m_{\tilde{g}} = 600$ GeV and a neutralino mass $m_{\tilde{\chi}_{1}^{0}} = 100$ GeV 
(the on-shell case is similar).

Without any ISR, we are left with $4$ jets in the final state.  
In order to do any proper analysis, it needs to be determined which of the $3$ possible pairings of jets is the correct combination. 
In Ref.~\cite{Rajaraman:2010hy}, the authors explored the possibility of extracting the correct jet combination 
using the invariant mass and transverse momentum of each diquark system. They noticed two prominent features. 
First, they noted the kinematic edge in the diquark invariant mass distribution, which is especially visible for combinations with high $p_T$. 
The excess of diquark combinations with an invariant mass larger than the kinematic edge value 
($m_{\tilde{g}}-m_{\tilde{\chi}_{1}^{0}}=500$ GeV) must all be incorrect combinations.
Second, those combinations which have invariant mass larger than the kinematic edge 
(which are all incorrect combinations) tail off quickly towards larger diquark $p_T$. 
Events below the edge (which are a mixture of correct and incorrect combinations) extend to higher $p_T$.  
Therefore, for the events with larger $p_T$, the ratio of correct to incorrect combinations increases.
%
%
%
\begin{figure}[th]
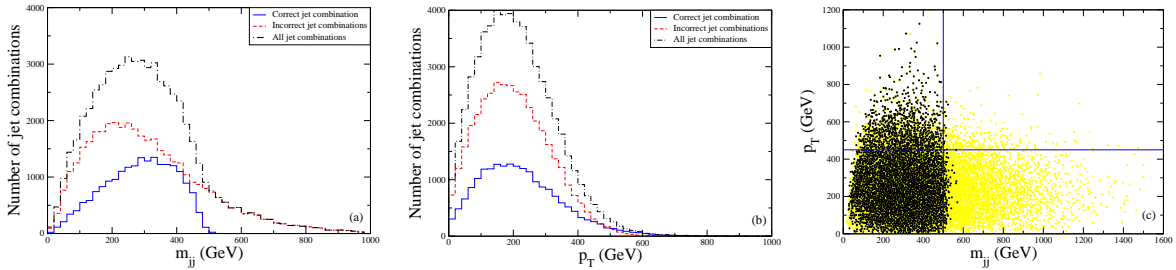

\begin{center}
\includegraphics*[angle=0,width=0.32\textwidth]{FIGURES/gluino_inv_new.eps}~~
\includegraphics*[angle=0,width=0.32\textwidth]{FIGURES/gluino_pT_new.eps}~
\includegraphics*[angle=0,width=0.32\textwidth]{FIGURES/pt_invmass_new.eps} 
\caption{The invariant mass (a) and $p_{T}$ distributions (b) of the jet combinations in the final state of the gluino pair production. 
(c) Diquark transverse momentum versus invariant mass for the correct (in black) and incorrect (in yellow) jet combinations. 
The vertical and horizontal lines represent the cuts used to determine the correct jet combination. 
These distribution show the motivation for the cuts proposed in~\cite{Rajaraman:2010hy}. }
\label{fig:gluino_pt_inv}
\end{center}
\end{figure}

To reproduce the results of Ref.~\cite{Rajaraman:2010hy} we use the MadGraph/MadEvent package~\cite{Alwall:2011uj} to generate a total of $10$K events at the LHC with a center of mass energy of $7$ TeV. 
Simulating the effects caused by detector resolution in the hadronic calorimeter, the $4$-momenta of the final state jets are smeared via
\begin{equation}
\left(\frac{\delta E}{E}\right)^{2} = \frac{a^{2}}{E} + b^{2}  \, ,
\label{eq:LHCres}
\end{equation}
where $a = 0.5~(0.1)~\sqrt{\text{GeV}}$ and $b = 0.03~(0.007)$ for jets (leptons)~\cite{RichterWas:2009wx,Bayatian:2006zz,Ball:2007zza}.

Fig.~\ref{fig:gluino_pt_inv} shows (a) the $p_{T}$ and (b) invariant mass distributions for the correct (blue, solid), incorrect (red, dashed) and all (black, dot-dashed) combinations of jets. 
The different normalizations of the three histograms are understood since there are two incorrect pairs and one correct pair in each event. 
In the invariant mass distribution of the correct jet combination there is a clear cutoff at $m_{\tilde{g}}-m_{\tilde{\chi}_{1}^{0}}=500$ GeV. 
A small fraction of events beyond this kinematic edge occur due to detector resolution. 
In the $p_{T}$ distribution at high value ($p_{T} \gtrsim$ 500 GeV) the ratio of correct to incorrect jet combinations becomes $\gtrsim 1$.
The $p_{T}$ cut is chosen such that only $5\%$ of the combinations with an invariant mass above the kinematic edge (which are likely to be incorrect combinations) survive. 
As suggested, a cut on combinations with the invariant mass below the kinematic edge and above the $p_T$ cut can guarantee an event sample 
dominated by correct combinations. 
To extract the correct jet combination the following cuts are made:
\begin{itemize}
\item{The invariant mass of the diquark pair is less than $500$ GeV.}
\item{The $p_T$ of the diquark pair is greater than $450$ GeV.}
\end{itemize}
If only one combination of jets passes these cuts then the event is accepted, otherwise the event is discarded.  
Fig.~\ref{fig:gluino_pt_inv}(c) shows the diquark transverse momentum versus invariant mass of 
the correct (black) and incorrect (yellow) jet combinations. 
The vertical and horizontal lines represent expected cuts to reduce the number of incorrect combinations. 
Counting those events that pass two cuts ({\it i.e.,} the number of events in the left-upper corner of Fig.~\ref{fig:gluino_pt_inv}(c)), 
we find an event efficiency of $3.1\%$ with an event purity of $91.7\%$. 
This is consistent with the results in Ref.~\cite{Rajaraman:2010hy}.
Furthermore, a higher efficiency can be achieved at the cost of lower purity by imposing a lower $p_T$ cut. 
It is also shown in~\cite{Rajaraman:2010hy} that this method performs better than the hemisphere method.


\section{Improved method}
\label{sec:gg}

We note that the method discussed in the previous section has a low efficiency while the corresponding purity is relatively high. 
This is due to the long tail in the $p_T$ distribution of the diquark system. 
In general, for a large mass splitting between the mother and daughter particles, 
the $p_T$ of the resulting diquark system is relatively large.  Conversely, it is smaller for a narrower spectrum, {\it i.e.,}
the peak position in the $p_T$ distribution is shifted to a lower value in a narrower mass spectrum.
Therefore there is a correlation between the location of the peak in the distribution and the mass difference.
However the amount of shift (or the location of the peak) is not well enough predicted quantitatively for this observation to be useful, 
since the diquark system inherits some kinematic information of the mother particles at the production level.
An ISR jet is also problematic since the mother particle system will be boosted in the opposite direction of the ISR resulting in modification of the $p_T$ spectra. 
Unlike the invariant mass distribution, the $p_T$ distribution does not exhibit any kinematic edge. 
The distribution of the background (incorrect combinations) has a similar shape and, 
as argued, it decreases faster than the distribution of the correct combination above a certain value 
(Fig. \ref{fig:gluino_pt_inv}(b)). On the other hand, the invariant mass distribution shows a clear kinematic edge 
in the distribution of the correct combination but not in that of the incorrect combination, making the two distributions distinct (Fig. \ref{fig:gluino_pt_inv}(a)). 

One can recall a similar situation in the leptonic decay of the $W$ boson at hadron colliders. 
The mass of the $W$ boson in principle can be measured from the $p_T$ distribution of a decay product (a lepton).
It shows a nice kinematic edge at half of the $W$ mass when the $W$ boson is produced at rest. 
However, in reality the transverse mass is used instead. 
It shows a kinematic edge at the mass of the $W$ and the value of the end point is stable under the presence of initial state radiation. 
What variable would do such a job for the example in the previous section while increasing the event efficiency? 
In this paper, instead of cutting on $p_T$, we propose to use $M_{T2}$ as the second cut in addition to the invariant mass.
It is the most natural extension of the transverse mass in the final state when more than one missing particle is present. 
As advertised, it shows a clear kinematic edge and is stable ({\it i.e.,} keeping the end point at the same value) under a boost of the whole system. 
One should investigate how well it performs in terms of the event efficiency and purity. 

We consider the same process ($4$ jets and $2$ neutralinos) as in the previous section 
without any ISR and discuss a method to improve the event efficiency and purity. 
Then, we discuss how our results change with different mass spectra and the effect of uncertainty in mass measurements. 
Finally, we look at the same process with ISR and discuss how our results change if we are correctly able to identify the ISR jet.
First, we begin this section with a review of $M_{T2}$.

 \subsection{Brief review of $M_{T2}$}
 \label{sec:review}

The traditional $M_{T2}$ variable is defined as follows \cite{Lester:1999tx,Barr:2003rg}. 
Consider a pair production of a mother particle, $P$, and their decays down to a daughter particle of mass $M_d$.
For any given event, one can construct the transverse mass, $M_{iT}$, of each parent:
\begin{equation}
M_{iT}^2\equiv m_i^2+M_d^2+2(E_{iT} E_{iT}^d-\vec{p}_{iT}\cdot\vec{p}^{\, d}_{iT})\, ,
\label{MTparent}
\end{equation}
where 
\begin{equation}
E_{iT}   \equiv \sqrt{m_i^2+|\vec p_{iT}|^2}, \quad
E_{iT}^d \equiv \sqrt{M_d^2+|\vec p_{iT}^{\,d}|^2},
\label{E}
\end{equation}
is the transverse energy of the visible particles, $V_i$, and daughter particle, $d$, in the decay branch of each mother particle, correspondingly. 
$\vec p_{iT}$ and $m_i$ are the transverse momentum and invariant mass of the visible particles in branch $i$ ($i=1,2$).
The individual momenta, $\vec p_{iT}^{\,d}$, of the missing daughter particles, $d$, are unknown, 
but they are constrained by the measured missing transverse momentum, $\mptvec$, in the event:
\begin{equation}
\vec p_{1T}^{\,d} + \vec p_{2T}^{\,d} = \mptvec
\equiv - \vec{P}_T - \vec p_{1T}- \vec p_{2T}. 
\label{PTmiss}
\end{equation}
Here $\vec{P}_T$ is the Upstream Transverse Momentum (UTM), which is the transverse momentum of all other visible particles that are not considered in the decay chains. 
For the true values of the missing momenta, $\vec p_{iT}^{\,d}$, each transverse mass in Eq. \ref{MTparent} is bounded above by the true parent mass, $M_P$. 

This fact can be used in a rather ingenious way to define the Cambridge $M_{T2}$ variable \cite{Lester:1999tx,Barr:2003rg}. 
One takes the larger of the two quantities in Eq. \ref{MTparent} and minimizes it over all possible partitions of the unknown daughter momenta, 
$\vec p_{iT}^{\,d}$, subject to the constraint in Eq. \ref{PTmiss}:
\begin{equation}
M_{T2}
(\tilde M_d, \vec{P}_T, \vec{p}_{i T}) 
\equiv 
\min_{\vec p_{1T}^{\,d} + \vec p_{2T}^{\,d} = \mptvec }
\left\{\max\left\{M_{1T},M_{2T}\right\} \right\}\ .
\label{eq:mt2def}
\end{equation}
Since the masses of the mother and daughter particles are not known in advance, one should treat
$M_{T2}$ as a function of the trial mass ($\tilde M_d$) of the daughter particle.
For a given $P_T$, the endpoint $M_{T2}^{max}$ of this distribution 
gives the parent mass, $\tilde M_P$, as a function of the input 
trial daughter mass, $\tilde M_d$. 
An important property of $M_{T2}$ is 
\begin{equation}
\tilde{M}_P ( \tilde{M}_d = M_d) = M_{T2}^{max} ( \tilde{M}_d = M_d) = M_P \, .
\end{equation}
Then one is able to obtain a one dimensional relation between two masses.
In cases with non-zero UTM and composite visible system, a kink structure appears, allowing 
simultaneous determination of both masses (see \cite{Barr:2010zj} and references therein). 

In the case of $\vec{P}_T = 0$, the analytic expression for $M_{T2}$ can be obtained, 
while for $\vec{P}_T \neq 0$ case there is no analytic solution to minimization and 
one has to rely on a numerical code for computation.
A clever trick to get around this has been proposed in Ref. \cite{Konar:2009wn}. 
The idea is to project further the transverse quantities with respect to the non-zero UTM, 
in which case the momentum conservation in the direction perpendicular to the UTM is independent of the UTM (by construction)
and one is able to use an analytic expression for the computation of projected $M_{T2}$.
$M_{T2}$ has been extended to cases with non-identical mothers \cite{Barr:2009jv} and non-identical daughters \cite{Konar:2009qr}.
In principle, one can use it for an associated production as well as processes with two different daughter particles. 

\subsection{Case without ISR}

\begin{figure}[t]
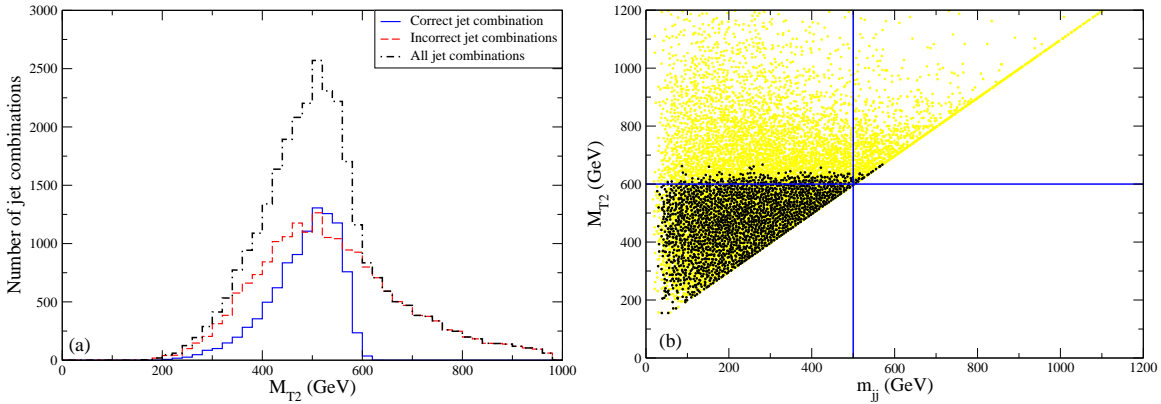

\begin{center}
\includegraphics*[angle=0,width=0.49\textwidth]{FIGURES/gluino_mT2_new.eps}
\includegraphics*[angle=0,width=0.49\textwidth]{FIGURES/mt2_invmass_new.eps}
\caption{(a) The $M_{T2}$ distribution of the correct and incorrect combinations of jets in the gluino pair production.  
For the correct jet combination a clear cutoff appears at the gluino mass. 
(b) The $M_{T2}$ versus invariant mass for the correct (in black) and incorrect (in yellow) jet combinations.  The vertical and horizontal lines represent the cuts used to determine the correct jet combination. }
\label{fig:gluino_mt2_inv}
\end{center}
\end{figure}

As an alternative method, instead of using transverse momentum, one uses $M_{T2}$ to help determine the correct jet combination.  
Fig.~\ref{fig:gluino_mt2_inv}(a) shows the $M_{T2}$ distribution for the correct (blue, solid), incorrect (red, dashed) and all (black, dot-dashed) combinations of jets. 
Note that the normalization of the distribution of incorrect (all) pairs is twice (three times) higher than that of the correct distribution. 
The distribution for the correct jet combination has a clear cutoff at the gluino mass ($m_{\tilde{g}}=600$ GeV). 
We apply the same cut that was used in the $p_T$ versus $M$ method for the invariant mass of the jet combinations 
but now we also require require $M_{T2} < 600$ GeV. 
As before, we keep the events where only one combination of jets passes both cuts.  
Fig.~\ref{fig:gluino_mt2_inv}(b) shows the correct (black) and incorrect combinations (yellow) in $M_{T2}$ versus invariant mass. 
For events where only one combination of jets pass the two cuts, 
we obtain an event efficiency of $20\%$ with an event purity of $95\%$ as shown in Table~\ref{tab:mt2_combos}. 
Thus we have a concrete example showing that an appropriate choice of kinematic variable cuts enhances both the
efficiency and purity of our signal.

Unlike the cuts using $p_{T}$, a very small fraction of events with correct combinations are ruled out completely by our cuts 
({\it i.e.}, many events have $2$ or $3$ combinations that pass the $M_{T2}$ and invariant mass cuts). 
Table~\ref{tab:mt2_combos} shows the number of events passing our cuts broken down by the number of combinations that pass.  
With these cuts only $2\%$ of the correct combinations are discarded and 98\% of correct combinations survive, 
so it is possible to use more refined cuts to extract the correct combination from the events with $2$ and $3$ combinations 
which could potentially increase the event efficiency even above $20\%$.
In principle, one can apply a $p_{T}$ cut at this point. 
However, as argued, $p_{T}$ is dependent on the mass spectrum and ISR, and 
perhaps it should be the last variable to make a cut on.
Instead, we further increase the efficiency and purity of this procedure by simply taking a look at the events where only two combinations pass the cuts so far.  

\begin{figure}[t]
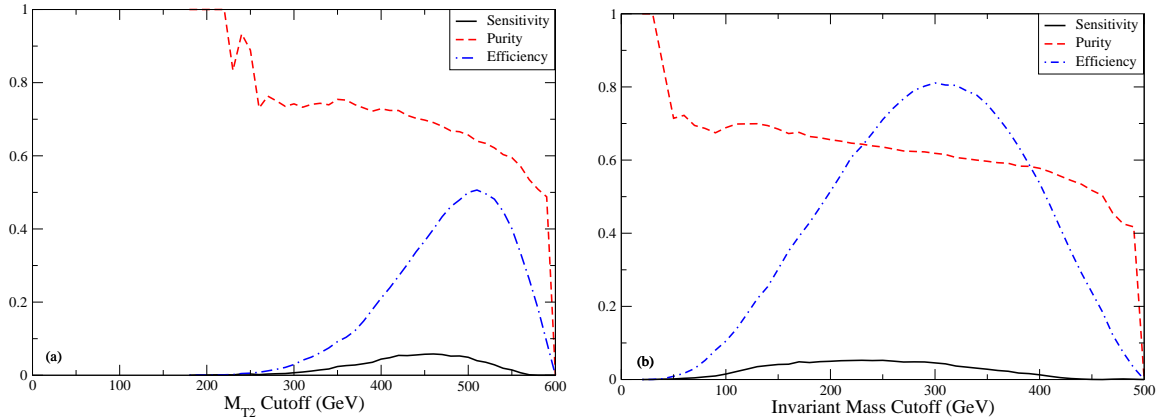
    
\begin{center}   
\centerline{
\includegraphics*[angle=0,width=0.48\textwidth]{FIGURES/mt2_2ndcut.eps}~~
\includegraphics*[angle=0,width=0.48\textwidth]{FIGURES/inv_2ndcut.eps}
}
\caption{The sensitivity (solid, black), efficiency (blue, dot-dashed) and purity (red, dashed) 
as functions of the (a) $M_{T2}$ and (b) invariant mass cutoff. 
The cutoffs are used to pick the correct combination out of the events 
where only two combinations passed the original cuts. We see that a $M_{T2}$ cut (invariant mass cut) of $460$ GeV ($230$ GeV) 
produces the largest sensitivity. 
Combining the results of the first set of cuts we have an overall efficiency of $33\%$ and a purity of $84\%$ ($41\%$ and a purity of $79\%$) 
with additional $M_{T2}$ cut (invariant mass cut). }
\label{fig:2ndcut}
\end{center}
\end{figure}

Making further cuts allows us to extract the correct combination from the two that are left over.  We discuss two such ways of extracting the correct combination using the invariant mass and $M_{T2}$ distributions.
\TABLE[ht]{
\centerline{
\begin{tabular}{|c|c|c|}	
\hline
Number of passed & Number of events & Number of events \\
combinations & & with correct combination \\
\hline
$0$ & $70$ & $-$ \\ 
$1$ & $2003$ & $1896$ \\
$2$ & $3135$ & $3076$ \\
$3$ & $4792$ & $4792$ \\
\hline
\end{tabular}
\caption{The breakdown of the number of events that have $0-3$ combinations that pass the $M_{T2}$ and invariant mass cuts.  The events with one correct combination give us a $20\%~(2003/10000)$ event efficiency with an event purity of $95\%~(1896/2003)$.  With these cuts only $2\%$ of the correct combinations are discarded so it is possible to use more refined cuts to extract the correct combination from the events with $2$ and $3$ combinations.}
\label{tab:mt2_combos}
} }
In Figs. \ref{fig:gluino_pt_inv}(a) and \ref{fig:gluino_mt2_inv}(a) we can see that for invariant masses and $M_{T2}$ that are below the kinematic cutoff the correct combination tends to be closer to the cutoff than the incorrect combination.  
We can take advantage of this by introducing a second cutoff.  
In the case of the $M_{T2}$ distribution we require that one combination have an $M_{T2}$ above this cutoff and the other is below.  
We then take the combination that is above this cutoff as the correct one.  
To select the cutoff we maximize the sensitivity $\epsilon D^2$, where the $\epsilon$ is the efficiency, $D= 2P-1$ is the dilution, and $P$ is the purity. 
This is a commonly used method for optimization in such situations \cite{YenChu}.
Fig.~\ref{fig:2ndcut}(a) shows that the maximum sensitivity occurs at an $M_{T2}$ cutoff at $460$ GeV 
which would give us an overall combined efficiency of $33\%$ and purity of $84\%$.
In the case of the invariant mass distribution, each combination has a pair of invariant masses so the cuts need to be modified slightly. 
From the two jet combinations we take the pair of invariant masses that have the largest separation between them.  
Placing a cut on the invariant mass we require that one invariant mass is above and the other is below the cutoff.  
We then take the combination with the invariant mass pair above the cutoff as the correct combination. 
Fig.~\ref{fig:2ndcut}(b) shows the scan of this invariant mass cutoff.  
We see that a cutoff of $230$ GeV gives the largest sensitivity to this set of events.  
Combined with the results from the first set of cuts we obtain a total event efficiency of $41\%$ and a purity of $79\%$.
%
%
%

\subsection{Uncertainty in particle masses}

Up to this point we have assumed the exact masses of the gluino and the lightest neutralino. 
What if the masses are not known with absolute certainty? 
To investigate the effects due to an uncertainty in mass measurements 
on the event efficiency and purity in finding the correct pairs, 
we generate events where both the masses of the gluino and neutralino fluctuate by $10\%$. 
Then using exactly the same cuts as before ($m_{jj} < 500$ GeV and $M_{T2} < $  600 GeV for all mass spectra) we compute the event efficiency and purity, 
as shown in Table~\ref{tab:mass_uncertainty}.

A $10\%$ increase in the gluino mass in general increases the invariant mass and $M_{T2}$ of the jet pairs, effectively reducing the probability that the jet pair will pass the cuts.  
In other words, some events with all three combinations surviving cuts end up having only one or two combinations filtered by this variation.  This results in an increasing event efficiency, but the contamination lowers the event purity.
On the other hand a gluino mass that is $10\%$ smaller decreases the invariant mass and $M_{T2}$ of the jet pairs, which increases the probability that a jet combination will pass all the cuts.  
This will lower the event efficiency but will raise the event purity due to the fact that in events with only one combination passing all the cuts, that combination is more likely to be the correct one.

The uncertainty in the neutralino mass produces the opposite effect on the event efficiency and purity.  
Because the neutralino is a daughter particle, increasing (decreasing) the neutralino mass will increase (decrease) the probability of a jet combination passing our cuts producing the opposite effect to the event efficiency and purity than a similar change in gluino mass.  However, it can be seen that the effect on the event efficiency and purity is much milder than that of the gluino due to the smaller change in neutralino mass.
Therefore, overestimating the mass of the mother particle (or underestimating the mass of daughter particle) decreases efficiency and increases purity, 
while underestimating the mother particle mass increases efficiency and decreases purity.
\TABLE[t]{
\centerline{
\begin{tabular}{|c||c|c|c|}
\hline 
   & $m_{\tilde{\chi}_1^0}=90$ & $m_{\tilde{\chi}_1^0}=100$ & $m_{\tilde{\chi}_1^0}=110$  \\
\hline \hline
$m_{\tilde{g}}$=660 & 0.31/0.61 & 0.30/0.63  & 0.28/0.65 \\
$m_{\tilde{g}}$=600 & 0.21/0.92 & 0.20/0.95  & 0.19/0.96 \\
$m_{\tilde{g}}$=540 & 0.13/1.00 & 0.12/1.00 & 0.11/1.00 \\
\hline
\end{tabular}
\caption{The event efficiency (left) and purity (right) with a $10\%$ uncertainty in the gluino and neutralino masses (in GeV). 
For all mass spectra the same cuts, $m_{jj} < 500$ GeV and $M_{T2} < $  600 GeV, are applied.}
\label{tab:mass_uncertainty}
} }

\subsection{Dependence on mass spectrum}

In principle, this method does not require a large mass splitting between the gluino and neutralino in order to be effective. 
To quantify this we scan the gluino and neutralino mass over a range of values. 
Assuming that the gluino and neutralino masses are known, we take the following cuts on the invariant mass and $M_{T2}$:
\begin{eqnarray}
m_{jj}   &<& m_{\tilde{g}}-m_{\tilde{\chi}^{0}_{1}} \, , \label{eq:cut1} \\
M_{T2} &<& m_{\tilde{g}} \, . \label{eq:cut2}
\end{eqnarray}
The results are summarized in Table~\ref{tab:purity}.

In the increasing $m_{\tilde{g}}$ direction, the invariant mass and $M_{T2}$ distributions of the correct and incorrect jet combinations 
becomes more spread apart.  In general this would increase the chance that a certain jet combination will be cut, reducing the event efficiency. 
However, the events that survive with only one passed combination will more likely be the correct combination, leading to a larger event purity. 
In the increasing $m_{\tilde{\chi}_{1}^{0}}$ direction, the invariant mass and $M_{T2}$ distributions of the correct and incorrect jet combinations becomes more compressed, 
resulting in higher efficiency and lower purity.
\TABLE[t]{
\centerline{
\begin{tabular}{|c||c|c|c|c|c|}
\hline 
   & $m_{\tilde{\chi}_1^0}=50$ & $m_{\tilde{\chi}_1^0}=100$ & $m_{\tilde{\chi}_1^0}=150$ & $m_{\tilde{\chi}_1^0}=200$ & $m_{\tilde{\chi}_1^0}=250$  \\
\hline \hline
$m_{\tilde{g}}$=700 & 0.15/0.94 & 0.17/0.94  & 0.20/0.93 & 0.22/0.93 & 0.28/0.93\\
$m_{\tilde{g}}$=600 & 0.17/0.95 & 0.20/0.94  & 0.24/0.93 & 0.26/0.93 & 0.28/0.92\\
$m_{\tilde{g}}$=500 & 0.21/0.94 & 0.24/0.93  & 0.28/0.93 & 0.31/0.93 & 0.36/0.91\\
$m_{\tilde{g}}$=400 & 0.25/0.94 & 0.30/0.93  & 0.34/0.91 & 0.37/0.91 & 0.43/0.87\\
$m_{\tilde{g}}$=300 & 0.30/0.93 & 0.36/0.92  & 0.42/0.90 & 0.48/0.85 & 0.54/0.80\\
\hline
\end{tabular}
\caption{The event efficiency (left) and purity (right) for several values of the gluino and neutralino masses (in GeV).  
For each mass spectrum the invariant mass and $M_{T2}$ cuts are given in Eqs. \ref{eq:cut1} and \ref{eq:cut2}.
}
\label{tab:purity}
}}
%
%
%

\subsection{Effect of ISR}

Since gluino pair production gives a multi-jet final state, it is important to address an issue with isolating ISR jets.
The ability to identify ISR is essential for using the discussed procedure: 
if the jets from ISR are not correctly isolated then it would be impossible to find the correct jet combination. 
Several different methods have been proposed to identify an ISR jet \cite{Alwall:2010cq,Alwall:2009zu,Nojiri:2010mk,Konar:2008ei, Konar:2010ma,Krohn:2011zp,Alwall:2007fs,Plehn:2005cq}. 
To analyze the gluino pair production with a single ISR jet we concentrate on the procedure introduced in Ref.~\cite{Alwall:2009zu}. 
In order to identify the ISR jet we use the following procedure.
\begin{enumerate}
\item{The jets with the highest momentum are not considered to be the ISR jet and are automatically placed in different gluino decay chains.}
\item{For the three other jets, each one is taken out and $M_{T2}$ is calculated from all the possible combinations of the remaining jets.  The minimum $M_{T2}$ of the possible jet combinations is identified as $M_{T2}^{(i)}$.}
\item{The jet associated with the minimum of the $M_{T2}^{(i)}$'s is taken to be the ISR jet.}
\end{enumerate}

Using this procedure alone we find that the ISR jet can be correctly identified $24\%$ of the time.  This number can be improved slightly by putting a minimum cutoff for the $M_{T2}$ of min$(M_{T2}) > 500$ GeV.  In general, if a correct jet is taken out then the resulting $M_{T2}$ becomes much smaller so this minimum $M_{T2}$ cutoff is designed to prevent the small $M_{T2}$ that arises from taking out a hard jet.  With this minimum cut included we find that the correct ISR jet can be identified $36\%$ of the time.  Both of these numbers are consistent with the results given in Ref.~\cite{Alwall:2009zu}.
If we look at the events that have correctly identified the ISR jet and apply the same cuts as before to isolate the correct combination of jets, 
we can see how our procedure fares with the addition of ISR.  The results are given in Table~\ref{tab:mt2_combos_isr}.
We find that for the events where the ISR was identified without the $M_{T2}$ minimum cutoff, 
the efficiency of finding the correct combination is $19\%$ with a purity of $92\%$. 
Including the minimum $M_{T2}$ cut the efficiency is found to be $16\%$ also with a purity of $92\%$. 
Though the efficiency drops slightly there are more events in this group where the correct ISR jet is identified.  
It is very important to note that these results were obtained from the events where the ISR jet was correctly identified.  
If we were to include the events where the incorrect ISR jet was identified then our purity will suffer. 
This exemplifies the importance of correctly identifying the ISR jet in this process.
\TABLE[t]{
\centerline{
\begin{tabular}{|c||c|c||c|c|}	
\hline
Number of          & Number of   & Number of          & Number of                & Number of events        \\
passed               &   events        & events                &  events                      & with correct     \\
  combinations   &                      &   with correct       &   with $M_{T2}$ cut     & combination \\
                           &                     &   combination      &                                    & and $M_{T2}$ cut\\
\hline
$0$ & $10$ & $ - $ & $17$ & $ - $ \\
$1$ & $454$ & $419$ & $580$ & $533$\\
$2$ & $773$ & $754$ & $1103$ & $1083$\\
$3$ & $1176$ & $1176$ & $1942$ & $1942$\\
\hline
total events & $2413$ &$-$ & $3642$ & $-$ \\ 
\hline
\end{tabular}
\caption{The breakdown of the number of events that have $0$-$3$ combinations that pass the $M_{T2}$ and invariant mass cuts. 
The correct ISR jet has been identified using the two methods explained in the text. 
The events with one correct combination give us an event efficiency of $19\%$ and $16\%$ 
for the events without and with the $M_{T2}$ cutoff respectively. 
Both sets of events have an event purity of $92\%$.}
\label{tab:mt2_combos_isr}
} }
%


\section{Application: the $b$-$\ell$ ambiguity in the $t \bar{t}$ dilepton channel}
\label{sec:ttbar}

In this section, we apply the same idea as in Section~\ref{sec:gg} to resolve the two fold ambiguity that 
arises in $t\bar{t}$ production in the dilepton channel at the Tevatron and the LHC. 
Since there are two leptons associated in this analysis, the corresponding background is already small.  
Requiring one or two b-tagged jets makes the background negligible. 
As an illustration, we estimated the cross sections for the signal and SM backgrounds at the 7 TeV LHC.
We have used $m_t=173$ GeV. 
Assuming the cross sections summarized in Table \ref{table:xsections}, 
we simulate the signal ($t\bar{t}$) and background events with PYTHIA \cite{Sjostrand:2006za} and 
further process them with PGS \cite{PGS} for detector effects.
We require two leptons, at least two jets, and missing transverse energy. 
The default cuts in each MC event generator are applied along with an extra dilepton invariant mass cut,
40 GeV $<$ $m_{\ell\ell}$ $<$ 200 GeV.
Table \ref{table:nevents} summarizes the number of signal (S) and background (B) 
events in each channel for zero, one, and two b-tagged jets.
The background cross sections are reduced significantly 
by requiring at least two leptons ($S0/B0 \sim 3.27$) and 
b-tagging further reduces the backgrounds 
($S1/B1 \sim 24$ for one b-tagged jet and $S2/B2 \sim 75$ for two b-tagged jets).
Although this conclusion with a naive estimation of the background cross section may be affected slightly due to systematic uncertainties and higher order corrections, 
we do not expect significantly different results in this two-lepton and one or two $b$-tagged final state. 
Therefore we will not address background issues in the rest of this section and will assume the background is negligible. 

For the analysis in the rest of this section, we generate 10K $t\bar{t}$ events using the MadGraph/MadEvent package at both the Tevatron and the 7 TeV LHC. 
We force the $W$-boson to decay leptonically into either an electron or muon. 
For events simulated at the LHC, detector resolution was simulated as in Eq. \ref{eq:LHCres}.
For the Tevatron, the detector resolution is simulated following the parametrization in PGS \cite{PGS}, 
\begin{eqnarray}
\frac{\delta E}{E} &=& \frac{0.8}{ \sqrt{E} }     ~~~ ~~~~~~~~~{\rm for~jets}\, , \\
\frac{\delta E}{E} &=& \frac{0.2}{ \sqrt{E}} +0.01    ~~~ {\rm for~leptons}\, .
\label{eq:Tevatronres}
\end{eqnarray}
We also assume that the ISR is correctly isolated.  This can be justified given that we require two $b$-tagged jets for our study on combinatorial issues. 
To separate the correct and incorrect combinations, we use the same method using the invariant mass of each $b$-$\ell$ pairing as well as the $M_{T2}$ variable. 
The distributions for the correct and incorrect combinations for $M_{T2}$ are shown in Fig. \ref{fig:mt2_ttbar} for both Tevatron (in (a)) and LHC (in (b)). 
The correct combinations have a cutoff at $m_{top}$, while the incorrect combinations have a tail extending beyond this cutoff.  
Fig. \ref{fig:invmass_ttbar} shows the distributions for $m_{b\ell}$, the invariant mass of the $b$-jet lepton pairings at the Tevatron (in (a)) and the LHC (in (b)). 
Both show a kinematic edge at $\sqrt{m_{top}^2 - m_W^2} \approx 153$ GeV.  

%
\TABLE[t]{
\centerline{
\begin{tabular}{|c||c|c|c|c|}
\hline 
Process                      & generator  & order            & final state         & cross section (pb)    \\
\hline \hline
$t\bar{t}$                   & THEORY \cite{Kidonakis:2011jg}    & NNLL resummation & $t\bar{t}$          & 163                           \\
$W^+$                        & FEWZ \cite{FEWZ}      & NNLO             & $\ell^+\nu$ + jets     & 16670              \\
$W^-$                        & FEWZ       & NNLO             & $\ell^-\bar{\nu}$ + jets     & 11379              \\
$Z/\gamma^\ast$               & MCFM       & NLO             & $\ell^+\ell^-$ + jets & 3000                     \\
$W^+W^-$                     & MCFM \cite{MCFM}       & NLO             & inclusive           & 43                 \\
$W^+ +Z/\gamma^\ast$            & MCFM       & NLO              & inclusive           & 11.8              \\
$W^- +Z/\gamma^\ast$            & MCFM       & NLO             & inclusive            & 6.4             \\
$Z/\gamma^\ast +Z/\gamma^\ast$  & MCFM       & NLO              & inclusive           & 5.9               \\
\hline
\end{tabular}
\label{table:xsections}
\caption{\sl Cross sections for 7 TeV $pp$ collisions. $\ell$ includes $e$ and $\mu$. 
$Z/\gamma^\ast+jets$ are computed for 40 GeV $< m_{\ell\ell} <$ 200 GeV with default cuts in MCFM.} 
} }
%
%
%
%
%
\TABLE[t]{
\begin{tabular}{|c||c|c|c||c|c|c||c|c|c|}
\hline 
                      & B0   & S0    & B0+S0  & B1  & S1   &  B1+S1 & B2 & S2   & B2 +S2   \\
\hline \hline
$e^+e^-$               & 608  & 990   & 1598  & 61  & 682   &  743  &  7 & 246  & 253 \\
$e^+\mu^- + e^-\mu^+$  & 97   & 2261  & 2358  & 6   & 1534  & 1540  & 0  & 509  & 509 \\
$\mu^+\mu^-$          & 663  & 1220  & 1883   & 59  & 822   & 881   & 7  & 290  & 297  \\
\hline 
total                 & 1368 & 4471  & 5839   & 126 & 3038 &  3164  & 14 & 1045 & 1059  \\
\hline
\end{tabular}
\label{table:nevents}
\caption{\sl The number of signal (S) and background (B) events in each channel.
The numbers (1 and 2) represent the number of b-tagged jets. 
Applied cuts are nominal reconstruction cuts at PGS level and dilepton invariant mass cut, 
40 GeV $< m_{\ell\ell} <$ 200 GeV.}
}

Specific cut values, $M_{T2} <$ 176 GeV and $m_{b\ell} <$ 156 GeV, are found by maximizing the sensitivity, defined as $S = \epsilon {\left( 2 P - 1 \right)}^2 $. 
Events are selected if only one of the combinations passes these cuts. 
The efficiency ($\epsilon$) determines the fraction of $t\bar{t}$ events passing this criterion and the purity ($P$) is defined as the fraction of 
lepton-jet pairs that are correctly determined. 
One can optimize these cuts differently, depending on whether higher purity (but lower efficiency) or higher efficiency (but lower purity) is desired.
Table \ref{tab:ttbar_results_LHC} (Table \ref{tab:ttbar_results_Tevatron}) shows the results at the LHC (Tevatron) broken down by the number of combinations 
that pass the invariant mass and $M_{T2}$ cuts in each event.  When the restriction that 
only one combination passes the cuts is applied, we obtain an efficiency of 51.7\% (39.9\%) and a purity of 94.9\% (91.9\%). 

The CDF collaboration uses several methods to correctly pair the $b$-jet and the lepton \cite{YenChu}: 
\mblmax, the kinematic method, and the neutrino phi weighting method. Let us explain each method briefly.

{\it \mblmax method}: 
Since there are two leptons and two $b$-jets in each event, there are two exclusive pairing options so there are 
in total four $m_{b\ell}^2$ values. One takes the pairing option having the maximum $m_{b\ell}^2$ as the incorrect one. 
As shown in Fig.~\ref{fig:invmass_ttbar}, above a certain value, larger $m_{b\ell}^2$ comes often from the incorrect combination.
To improve the purity of pairing, the events that have maximum $m_{b\ell}^2$ larger than a certain cut value are selected. 
A larger cut value improves the purity at the cost of reduced efficiency. 
In order to find the best cut value (leading to maximum sensitivity, $\epsilon (2P-1)^2$) on \mblmax, 
one can scan through various maximum invariant mass cut values and select the events 
having maximum invariant mass larger than the cut. The efficiency and purity are calculated for each cut value. 
\FIGURE[t]{
\centerline{ \hspace{0.3cm}
\epsfig{file=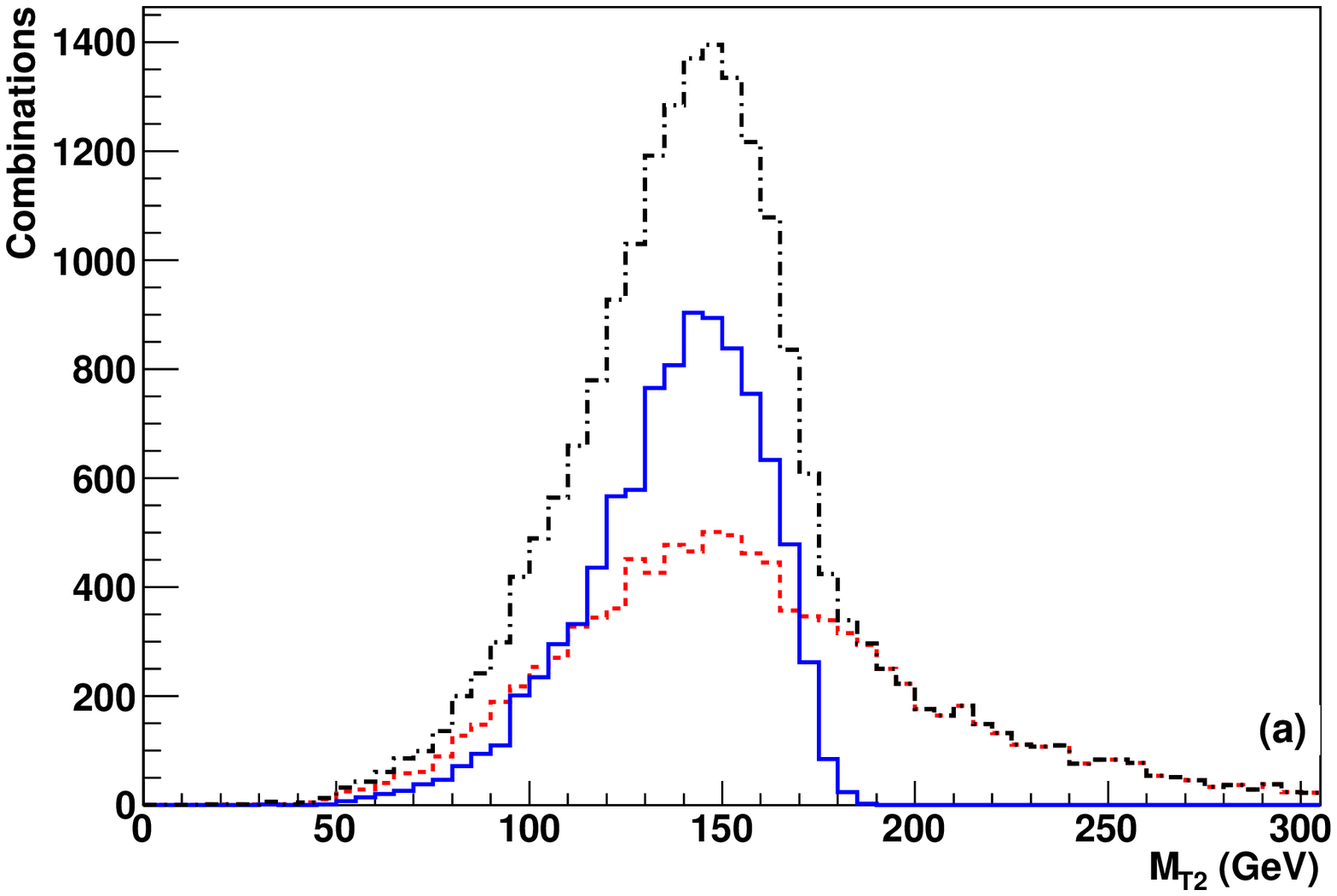, height=5.6cm} \hspace{-0.5cm}
\epsfig{file=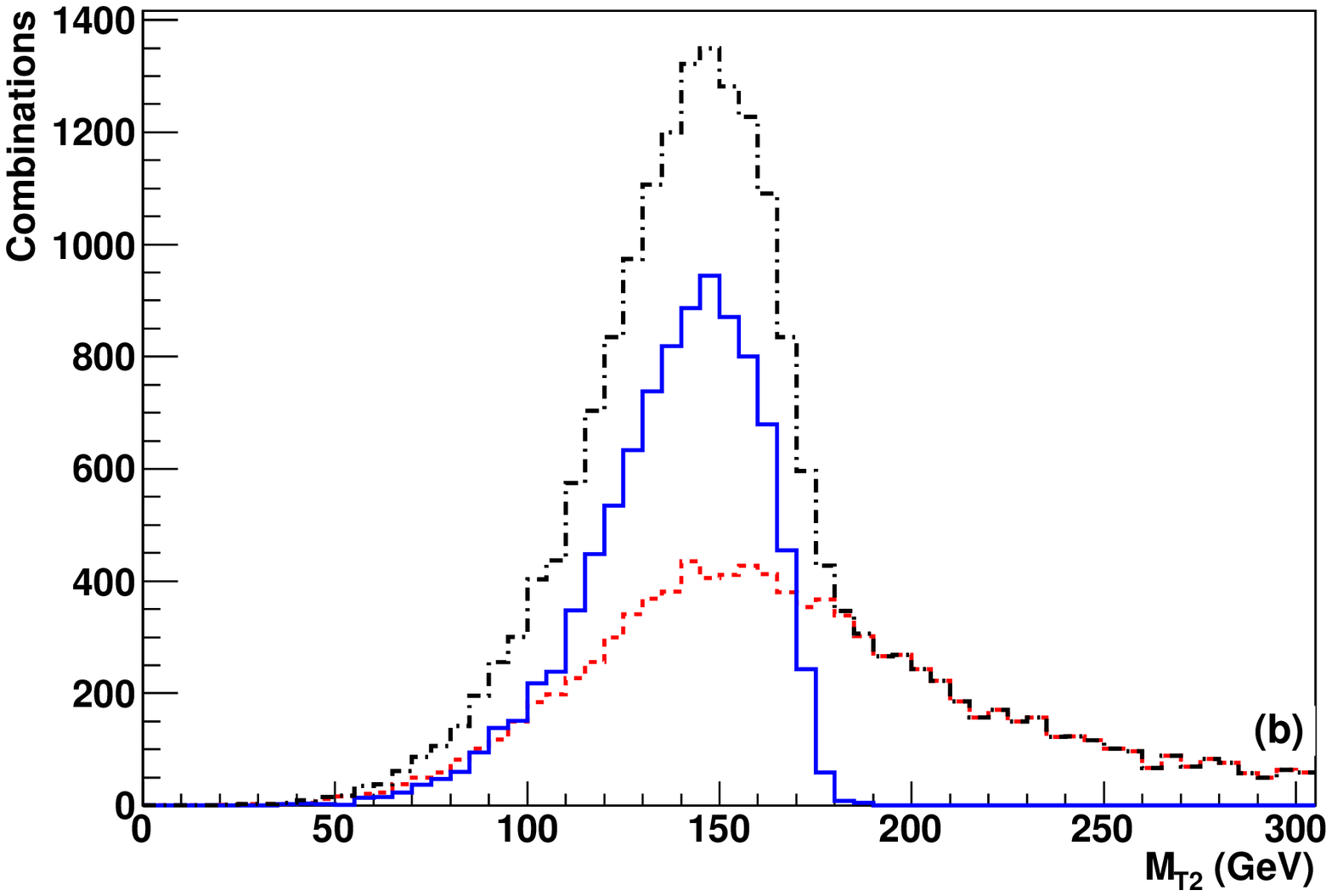, height=5.6cm}  }
\caption{The $M_{T2}$ distributions of the $b$-$\ell$ combinations in the final state of the $t \bar{t}$ pair production at the Tevatron (a) and the LHC (b). 
They show the clear kinematic edges at $m_{top}$. The distribution of correct combinations is shown in blue (solid), while the incorrect one is in red (dotted) 
and all combinations are in black (dot-dashed).}
\label{fig:mt2_ttbar}  }
\FIGURE[ht]{
\centerline{ \hspace{0.3cm}
\epsfig{file=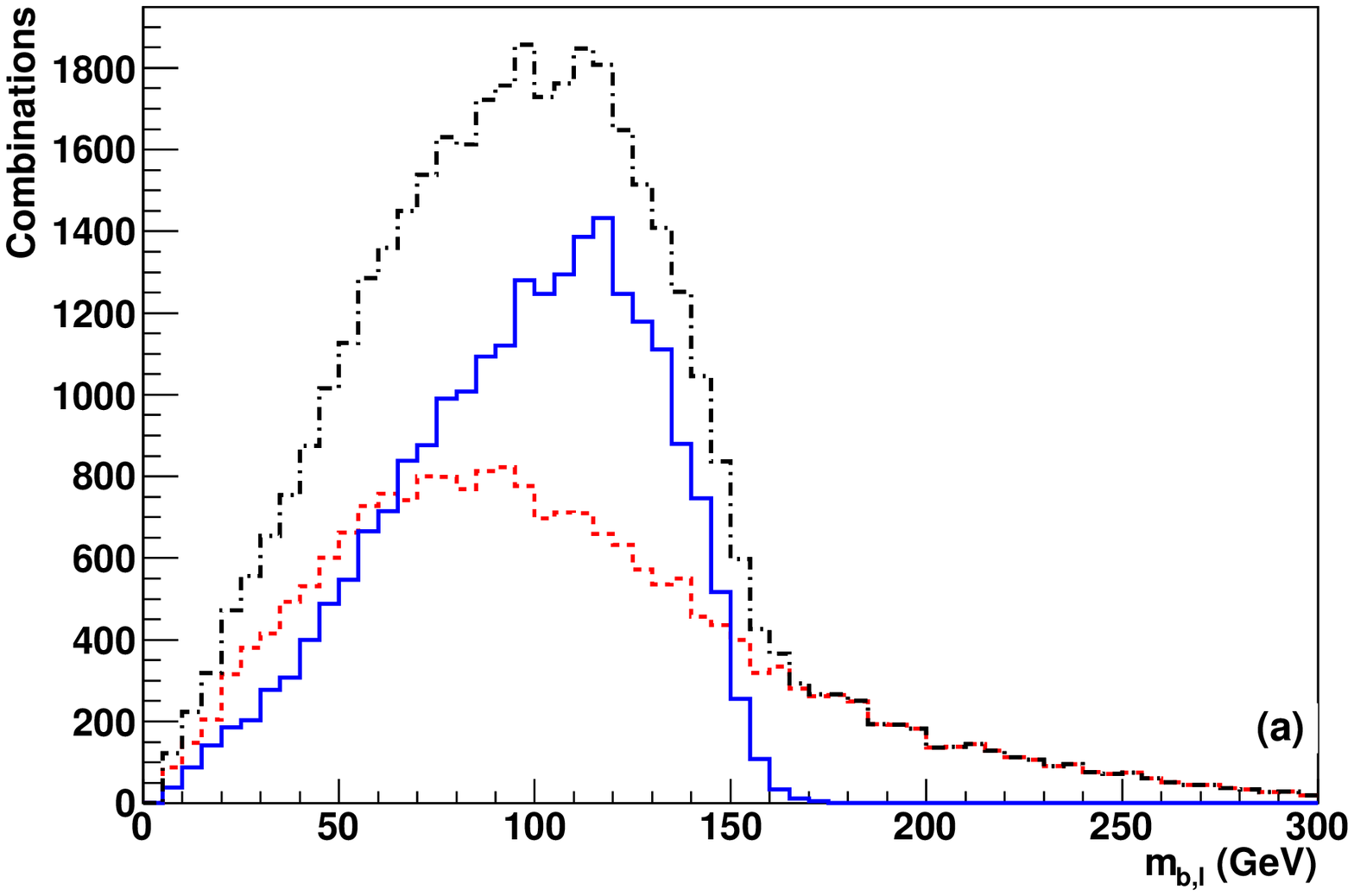,height=5.6cm} \hspace{-0.5cm}
\epsfig{file=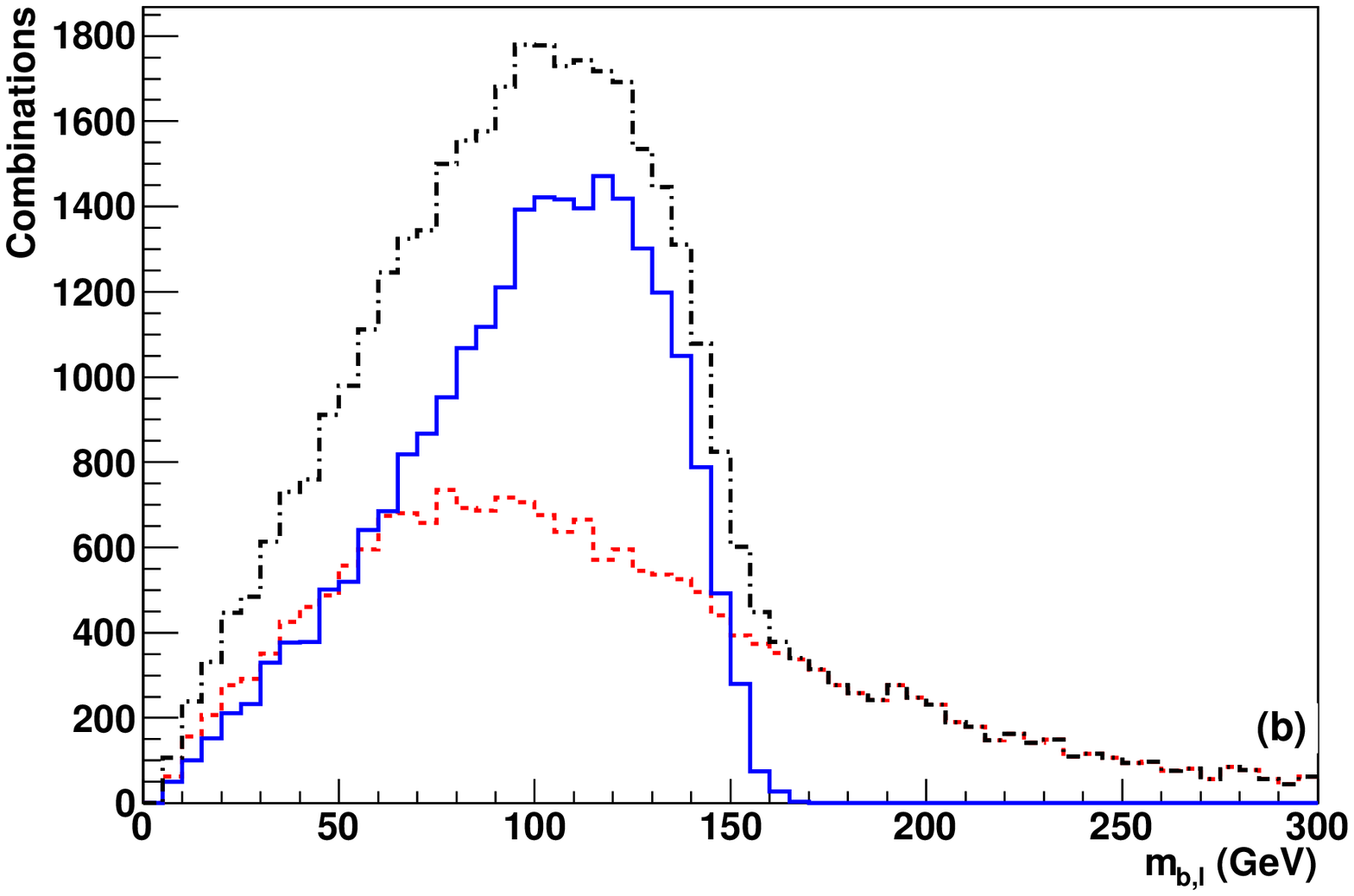,height=5.6cm} }
\caption{The invariant mass ($m_{b \ell}$) distributions of the $b$-$\ell$ combinations in the final state of the $t \bar{t}$ pair production at the Tevatron (a) and the LHC (b).  
They show the clear kinematic edges at $\sqrt{m_{top}^2 - m_W^2}$.}
\label{fig:invmass_ttbar}  }

{\it Kinematic method}: the kinematic method is a variation of the method used in the top mass analysis in the dilepton channel. 
This method is based on resolving the full energy-momentum conservation equation set for all the particles in the final sates. 
By a process of eliminating variables one is left with a 4th-order 
polynomial equation. Therefore up to 4 real solutions are possible. The CDF collaboration uses a numerical method to solve equations and look for up to two solutions.
If there are two solutions found, they choose the one having smaller $m_{t\bar{t}}$. 
A variation of this method has been used in studies of spin correlation of the $t\bar{t}$ system  \cite{CDF9824}.
The kinematic method has slightly higher efficiency than the \mblmax method 
but its purity is significantly lower \cite{YenChu}. 
Performance of the neutrino phi weighting method is similar to the kinematic method \cite{YenChu}.
The matrix element method can be also used but we do not find any comments on the pairing purity in the literature.
\TABLE[th]{
\centerline{
\begin{tabular}{|c|c|c|}	
\hline
Number of passed & Number of events & Number of events \\
 combinations & & with correct combinations \\
\hline
$0$ & $ 248 $   & $-$ \\
$1$ & $  5173$  & $4908$ \\
$2$ & $  4575$  & $ 4416$ \\
\hline
\end{tabular}
\caption{Results broken down by number of combinations passing the both the invariant mass and $M_{T2}$ cuts for events at the 7 TeV LHC. 
By selecting events with only one event passing, an efficiency of $51.7\%$ with a purity of $94.9\%$ is obtained. }
\label{tab:ttbar_results_LHC}
}}
\TABLE[ht]{
\centerline{
\begin{tabular}{|c|c|c|}	
\hline
Number of passed & Number of events & Number of events \\
 combinations & & with correct combinations \\
\hline
$0$ & $230$ & $-$ \\
$1$ & $3990$ & $3665$ \\
$2$ & $5778$ & $5580$ \\
\hline
\end{tabular}
\caption{The same as Table~\ref{tab:ttbar_results_LHC} but at the Tevatron. 
The corresponding efficiency and purity are $39.9\%$ and $91.9\%$.}
\label{tab:ttbar_results_Tevatron}
} }

As a comparison with our method, we have considered the \mblmax method 
(denoted as ($m_{b\ell}$, none) in Figs. \ref{fig:CDFsensitivity} and \ref{fig:combinedsensitivity}, and  
Tables \ref{tab:CDF-Tevatron} and \ref{tab:CDF-LHC}) used by the CDF collaboration. 
Fig. \ref{fig:CDFsensitivity} shows the purity-efficiency relation with the sensitivity at the Tevatron (in (a)) and the LHC (in (b)). 
This one-dimensional relation is due to a degree of freedom in choosing the invariant mass cut. 
The original CDF method is denoted as ($m_{b\ell}$, none) (solid, red) while 
($M_{T2}$, none) (dashed, black) represent the same scheme with $m_{b\ell}$ replaced by $M_{T2}$. 
At the Tevatron when a cut of $m_{b\ell} >$ 151 GeV is made, the maximum sensitivity is obtained and 
this method results in an efficiency of 37.2\% and a purity of 96.3\% for top quark production. 
The corresponding cuts for ($m_{b\ell}$, none) are are shown in Table  \ref{tab:CDF-Tevatron}.
These are comparable to results obtained with our proposal. 
For the LHC, both purity and efficiency are higher (see Fig. \ref{fig:CDFsensitivity}(b) and Table \ref{tab:CDF-LHC}).

While the initial requirement of only one combination passing the cuts for our method results in reasonable values for efficiency, 
we can still improve this by making use of the events in which both combinations pass the cuts (nearly 50 percent of the events for the LHC). 
It is possible to implement another reconstruction method on these events.
One example would be to combine our method with the \mblmax method to improve the total number of events in which a good combination can be selected, while not causing too large of an adverse effect on the purity. 
In this case, however, we do not find a noticeable improvement over the original method.

Instead, we take a different approach to increase the event efficiency and purity.
The CDF procedure cuts out events in which the combination with the highest $m_{b\ell}$ or $M_{T2}$ is not over the cut value, thus giving a lower efficiency.  We look at those events and apply a second cut on the variable not used in the first case, thus applying the method with $M_{T2}$ for events not passing the $m_{b\ell}$ selection, ($m_{b\ell}$, $M_{T2}$), and vice versa, ($M_{T2}$, $m_{b\ell}$). We find that the $M_{T2}$ and invariant mass together with the CDF scheme lead to much better efficiency and purity 
than those obtained with one of them only.
Fig. \ref{fig:combinedsensitivity} shows contours of the efficiency (red, dashed), purity (blue, dotted) and sensitivity (black, solid) 
in the $M_{T2}$-$m_{b\ell}$ plane at the Tevatron (a) and the LHC (b). 
The cuts that correspond to the maximum sensitivity are shown in Table \ref{tab:CDF-LHC}. 
Similar results are obtained for both ($m_{b\ell}$, $M_{T2}$) and ($M_{T2}$, $m_{b\ell}$) cases, and this shows the order of the two cuts is not crucial. 
With both kinematic cuts, we improve the efficiency by 20\% while keeping the purity at the same level or slightly higher. 
\FIGURE[ht]{
\centerline{ 
\epsfig{file=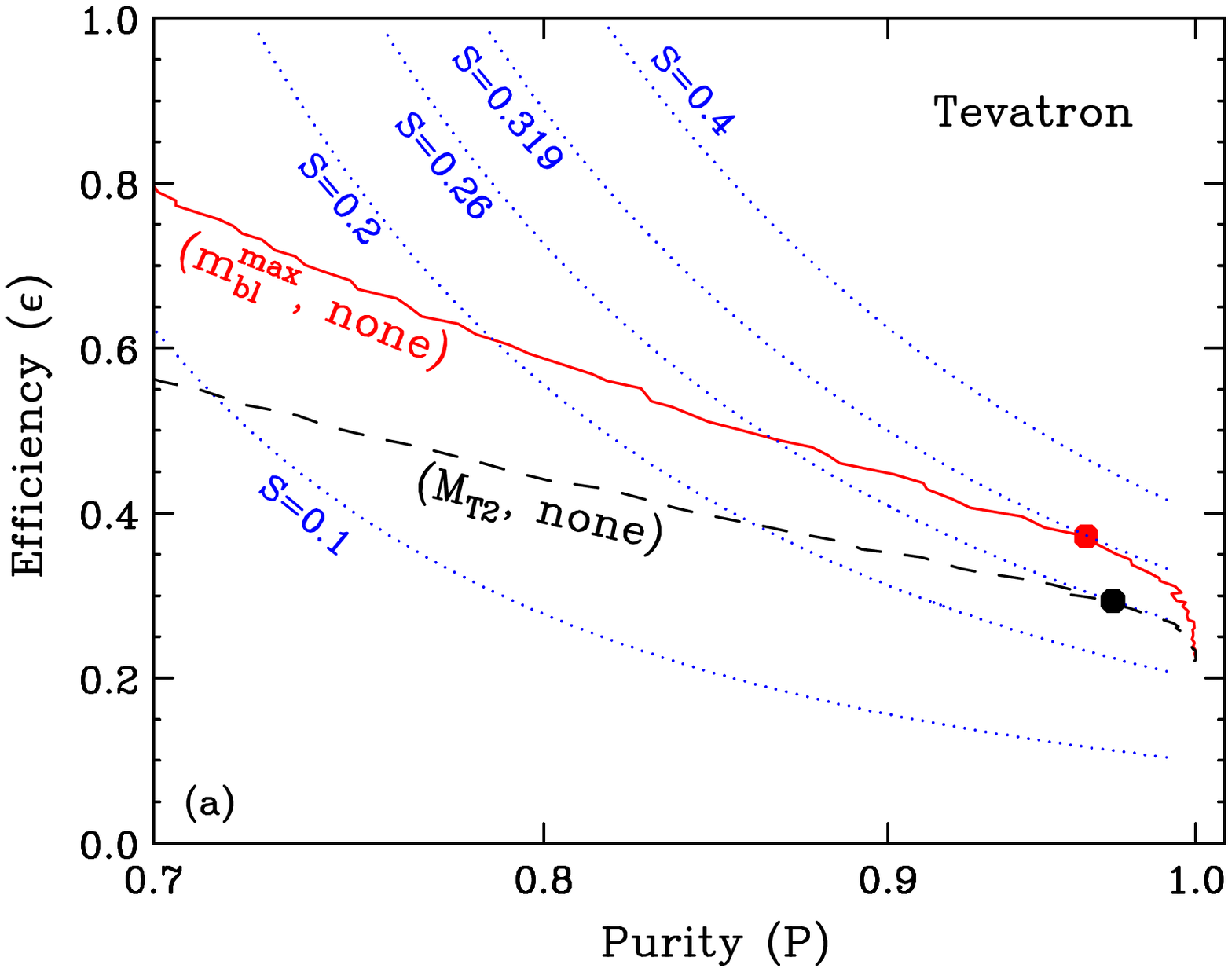,height=5.9cm}  \hspace{0.2cm}
\epsfig{file=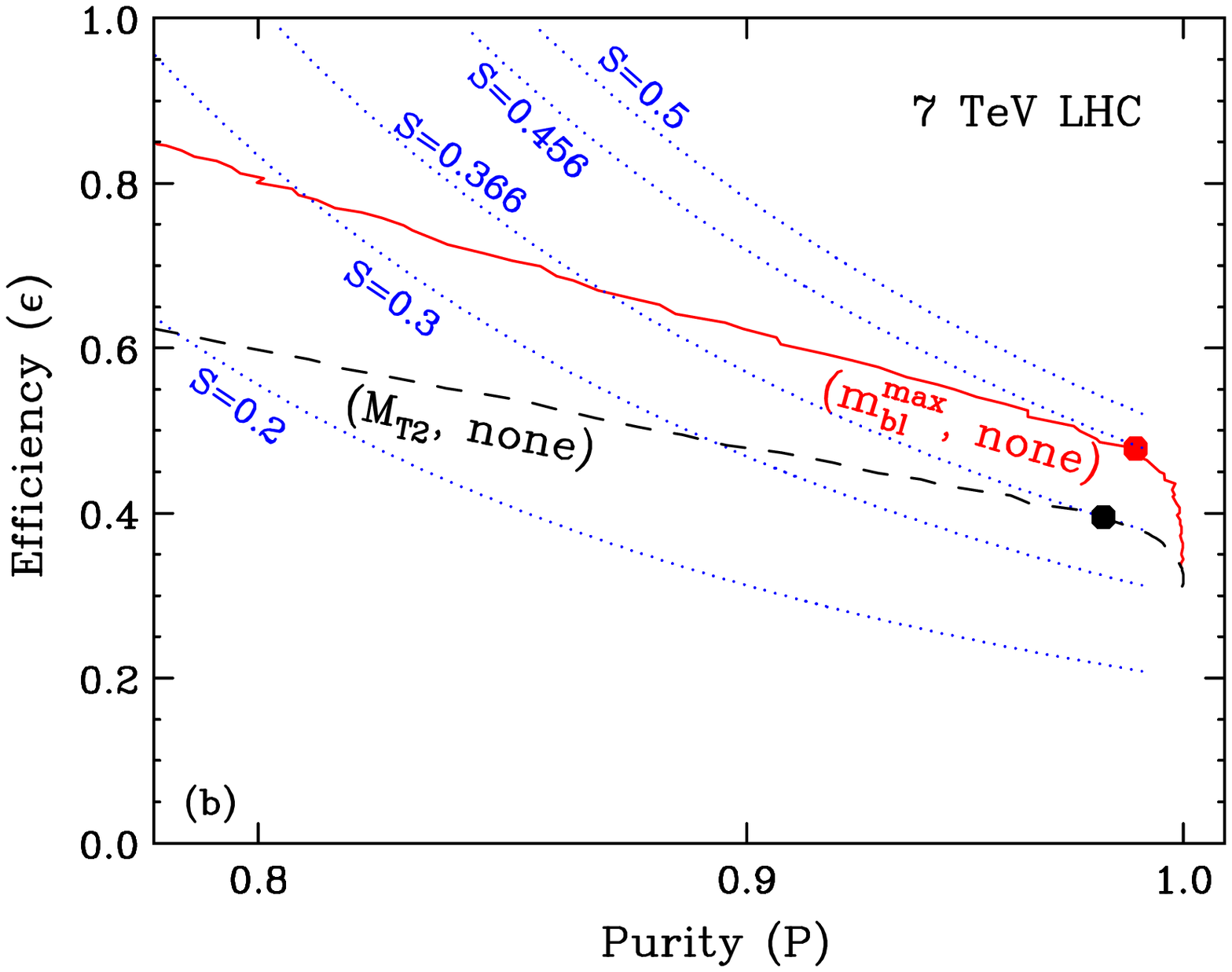,height=5.9cm} 
\caption{The efficiency, purity and sensitivity using the CDF method at the Tevatron (a) and the LHC (b). 
The original CDF method is denoted as ($m_{b\ell}$, none) (solid, red) while 
($M_{t2}$, none) (dashed, black) represent the same scheme with $m_{b\ell}$ replaced by $M_{T2}$. 
The contours of constant sensitivity are shown in dotted (blue) curves.
The dots represent the efficiency and purity that results in the maximum sensitivity, and 
the corresponding cuts are shown in Tables \ref{tab:CDF-Tevatron} and \ref{tab:CDF-LHC}.} 
\label{fig:CDFsensitivity}  } }
\vspace{0.1cm}
\FIGURE[ht]{
\centerline{ 
\epsfig{file=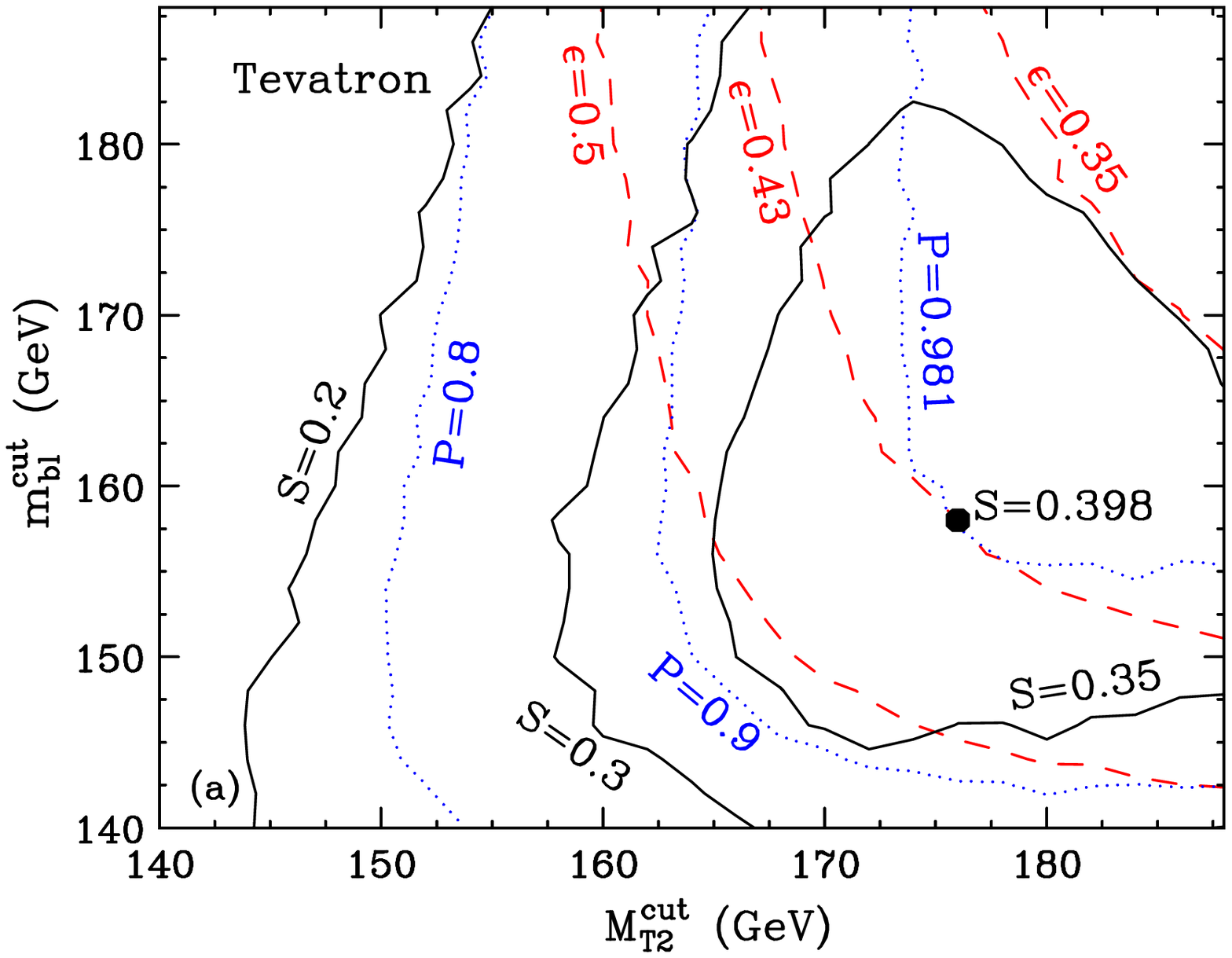,height=5.9cm}  \hspace{0.15cm}
\epsfig{file=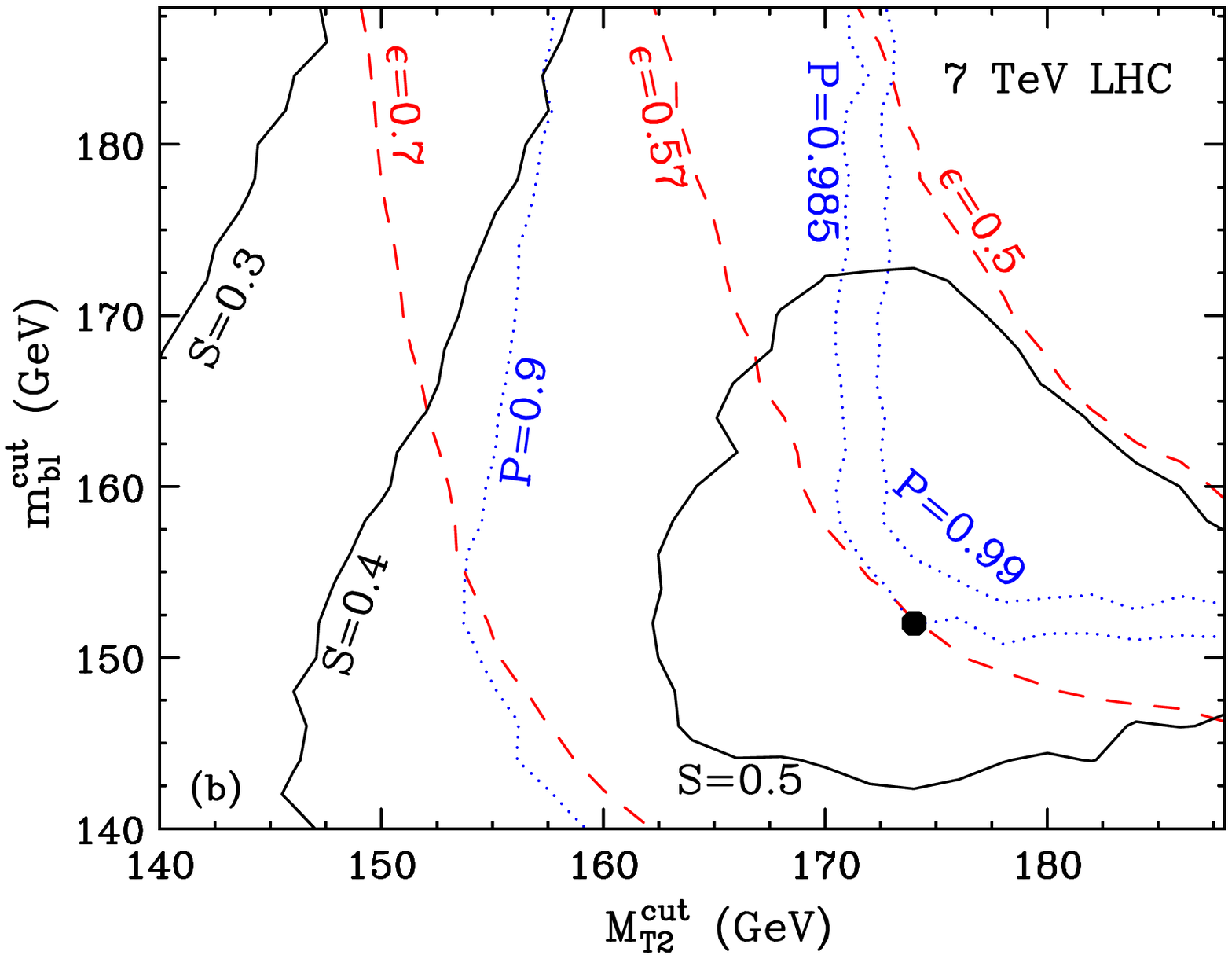,height=5.9cm} 
\caption{Contours of the efficiency (red, dashed), purity (blue, dotted) and sensitivity (black, solid) in the $M_{T2}$-$m_{b\ell}$ 
plane at the Tevatron (a) and the LHC (b). 
The dots represent the efficiency and purity that results in the maximum sensitivity. 
The corresponding cuts are shown in Tables \ref{tab:CDF-Tevatron} and \ref{tab:CDF-LHC}.} 
\label{fig:combinedsensitivity}  } }

While in this section we have discussed the combinatorial ambiguity in the $b$-$\ell$ pairings in the dilepton channel, 
the same method can be used in principle in the semi-leptonic channel as well.
In the latter case, often a $\chi^2$ (or a likelihood function) is defined with mass-shell constraints of 
the top quark and the $W$ boson on the both hadronic and leptonic sides, 
assuming the missing transverse momentum is solely due to the neutrino from the $W$ decay. 
In the procedure, from the on-shell condition of the leptonic $W$, 
an additional two-fold ambiguity in sign for the $z$-component (along the beam direction) of the neutrino momentum is necessarily introduced. 
The correct combination and sign are determined by minimizing the $\chi^2$ (or maximizing the likelihood function). 
The kinematic variables such as $M_{T2}$ and invariant mass should be able to help to increase the probability of finding the correct combination, 
when they are properly included in the $\chi^2$ or the likelihood functions.

We have assumed both $b$-jets are tagged but the proposed method here should apply in the final state with one $b$-tagged jet as well as 
in the dilepton channel with no $b$-tagged jets. 
In these cases, the candidates for the $b$-jet can be determined by different selection criteria. 
For instance, in the final state with one $b$-tagged jet, the hardest remaining jet could be a good candidate for the other $b$-jet while 
for the final state with no $b$-tagged jets, the two hardest jets are good choices. 
However the purity of the $t\bar{t}$ sample in the latter case will decrease significantly.
\TABLE[ht]{
\centerline{
\begin{tabular}{|c|c|c|c|c|}	
\hline
cut   & cut value (GeV) & efficiency ($\epsilon$) & purity ($P$) & sensitivity ($S$) \\
\hline
($m_{b\ell}$, none)     & (151, none)  & 0.372 & 0.963 & 0.319\\
($M_{T2}$, none)       &  (174, none)  & 0.294 & 0.972 & 0.261\\
($m_{b\ell}$, $M_{T2}$) &  (158, 176)  & 0.426 & 0.981 & 0.398\\
($M_{T2}$, $m_{b\ell}$) &  (175, 155)  & 0.442 & 0.971 & 0.392\\
\hline
\end{tabular}
\caption{Results at the Tevatron for invariant mass cut, $M_{T2}$ cut and the hybrid, following CDF scheme. 
The sensitivity is defined as $S=\epsilon (2P-1)^2$.}
\label{tab:CDF-Tevatron}
} }
\TABLE[ht]{
\centerline{
\begin{tabular}{|c|c|c|c|c|}	
\hline
cut   & cut value (GeV) & efficiency ($\epsilon$) & purity ($P$) & sensitivity ($S$) \\
\hline
($m_{b\ell}$, none)     & (152, none)  & 0.478 & 0.988 & 0.456 \\
($M_{T2}$, none)       &  (172, none)  & 0.395 & 0.981 & 0.366 \\
($m_{b\ell}$, $M_{T2}$) &  (152, 174)    & 0.574 & 0.985 & 0.539\\
($M_{T2}$, $m_{b\ell}$) &  (175, 155)    & 0.557 & 0.989 & 0.533 \\
\hline
\end{tabular}
\caption{The same as Table~\ref{tab:CDF-Tevatron} but for the LHC.}
\label{tab:CDF-LHC}
}}
%

\section{Summary and conclusions}
\label{sec:conclusions}

The $M_{T2}$ variable is originally introduced to measure masses of semi-invisibly decaying particles. 
It is the most natural extension to the transverse mass that was used for the $W$ discovery and mass measurement.
For the last few years, there have been many studies to determine masses, spins and couplings.
Most of them utilize the kinematics of events with missing energy.
The $M_{T2}$ variable is especially useful for relatively short decay chains where a traditional invariant mass method is limited.
The subsystem $M_{T2}$ allows for measurement of all masses in the decay chain, assuming combinatorial issues are sorted out.
It is also shown that the $M_{T2}$ variable shows an interesting kink structure due to the compositeness of the visible sector, 
the existence of ISR, or upstream transverse momentum from the heavier particle decays. 
It has been extended to include different mother/daughter particles masses. 
Finally it has been proposed as a potential solution for isolating ISR from signals with multiple jets. 

In this paper we have investigated the feasibility of the use of $M_{T2}$ to resolve combinatorial issues at hadron colliders. 
We concentrated on a 4-jets signal where 3 possible partitions exist.
We have compared the performance of this method with the $p_T$ versus invariant mass method and 
showed that efficiency increased up to a factor of 5 for the same purity of the sample.
One of the advantages of the kinematic methods is that there is little dependence on the mass spectrum (model dependence), 
unlike a choice of $p_{T}$. We have found that our results for efficiency and purity remain similar for different choices of mass parameters.
We showed that the kinematic variable suggested for mass determination is also useful for resolving combinatorial issues in signals, 
as well as for isolating ISR jets from signal jets.

A similar idea is applied to the $t\bar{t}$ dilepton system and we found that the obtained efficiency and 
purity are comparable to those using the current CDF method. 
A variation of their method with $M_{T2}$ results in improvement of the efficiency by 20\% at the Tevatron and 
25\% at the LHC, while the corresponding purity remains the same.
It is desirable and important for experimental collaborations to revisit the method with a full detector simulation.

\bigskip

\acknowledgments
We thank Yen-Chu Chen for helpful discussion and providing relevant references regarding $b$-$\ell$ ambiguity in the $t\bar{t}$ production.
This work is supported in part by US Department of Energy grants DE-FG02-04ER14308 and by National Science Foundation grants PHY-0653250. 
KK is supported partially by the National Science Foundation under Award No. EPS-0903806 and 
matching funds from the State of Kansas through the Kansas Technology Enterprise Corporation. 
KK acknowledges the hospitality of TASI-2011 at the University of Colorado, where some of this work was undertaken.




\end{document}